\documentclass[showpacs,preprintnumbers,amsmath,amssymb,prb,twocolumn,superscriptaddress,footinbib]{revtex4}
\usepackage{graphicx}
\usepackage{bm}

\newcommand{\bk}{{\boldsymbol{k}}}

\newcommand{\bp}{{\boldsymbol{p}}}
\newcommand{\cdg}{c^\dagger_{\bk \sigma}}
\newcommand{\cdgup}{c^\dagger_{\bk \uparrow}}
\newcommand{\cdgdn}{c^\dagger_{\bk \downarrow}}
\newcommand{\cu}{c_{\bk \sigma}}

\newcommand{\dud}{d_{\bp \sigma}}

\newcommand{\ex}[1]{\mathrm{e}^{#1}}
\newcommand{\im}{\mathrm{i}}

\newcommand{\du}{\mathrm{d}}

\newcommand{\epstd}[1]{\tilde{\varepsilon}_{#1,\bk}}

\newcommand{\cdgtd}[1]{\tilde{c}^\dagger_{#1,\bk}}
\newcommand{\ctd}[1]{\tilde{c}_{#1,\bk}}
\newcommand{\A}{\mathrm{A}}
\newcommand{\B}{\mathrm{B}}
\newcommand{\T}{\mathrm{T}}

\begin{document}


\title{Tunnelling between non-centrosymmetric superconductors with
  significant spin-orbit splitting studied theoretically within a
  two-band treatment}


\author{K. B{\o}rkje}
\author{A. Sudb{\o}}
\affiliation{Department of Physics, Norwegian University of Science
  and Technology, N-7491 Trondheim, Norway}


\date{\today}

\begin{abstract}
  Tunnelling between non-centrosymmetric superconductors with
  significant spin-orbit splitting is studied theoretically in a
  two-band treatment of the problem. We find that the critical
  Josephson current may be modulated by changing the relative angle
  between the vectors describing absence of inversion symmetry on each
  side of the junction. The presence of two gaps also results in
  multiple steps in the quasiparticle current-voltage characteristics.
  We argue that both these effects may help to determine the pairing
  states in materials like CePt$_3$Si, UIr and Cd$_2$Re$_2$O$_7$. We
  propose experimental tests of these ideas, including scanning
  tunnelling microscopy.
\end{abstract}


\maketitle


Superconductors where inversion symmetry is absent and spin-orbit
splitting is significant have recently attracted considerable
attention.
\cite{edelstein,gorkov,samokhin,frigeri,sergienko,fujimoto_hayashi,mineev,bonalde_yogi_izawa}
Much of this interest was initiated by the discovery of
superconductivity in CePt$_3$Si \cite{bauer} and UIr.\cite{akazawa}
In addition to not having an inversion center, band structure
calculations on CePt$_3$Si \cite{samokhin} have shown that spin-orbit
coupling splits otherwise degenerate bands by 50-200 meV near the
Fermi level. This is much larger than $k_\B T_{\mathrm{c}}$, where
$T_c$ is the critical temperature for superconductivity. This must be
taken into account when describing superconductivity in these systems.
These materials also order magnetically, which could influence the
nature of the superconducting state. However, at least for CePt$_3$Si,
there seems to be little communication between superconductivity and
magnetic order.\cite{amato}

Another superconductor of interest in this context is
Cd$_2$Re$_2$O$_7$. This material has a structural phase transition
where a center of symmetry is lost.\cite{sergienko2} When certain ions
in the unit cell are displaced throughout the lattice, internal
electric fields are induced, giving rise to spin-orbit splitting of
spin-degenerate states. Calculations and photoemission studies
\cite{eguchi} have indicated that this splitting will have a
significant influence on the electronic band structure. Thus,
Cd$_2$Re$_2$O$_7$ is similar to the materials mentioned above,
although simpler, since it shows no sign of magnetic
order.\cite{vyaselev_lumsden} Another pyrochlore superconductor that
might fall into this category is KOs$_2$O$_6$.\cite{shibauchi}

The spin-orbit splitting of otherwise degenerate bands demands a
two-band description of superconductivity in these materials. An
exotic feature of non-centrosymmetric superconductors with large
spin-orbit splitting is the possible absence of a definite parity of
the superconducting state.\cite{edelstein,gorkov,fujimoto_hayashi}
Experiments have indicated that CePt$_3$Si might be in such a pairing
state, a linear combination of spin-singlet and spin-triplet states,
and that the gap may contain line nodes.\cite{bonalde_yogi_izawa}
Cd$_2$Re$_2$O$_7$ and KOs$_2$O$_6$ seems to be nodeless,
however.\cite{vyaselev_lumsden,shibauchi}

In the present study, we will investigate tunnelling currents between
two superconductors where a two-band description is necessary in both
systems. Junctions involving one superconductor with spin-orbit split
bands have been studied in Refs. \onlinecite{sergienko} and
\onlinecite{yokoyama}. We will restrict ourselves to intraband Cooper
pairing without specifying the microscopic mechanism responsible for
this. See Refs. \onlinecite{brinkman_agterberg} for related work on
MgB$_2$ junctions.  Our main focus will be on non-centrosymmetric
superconductors with spin-orbit split bands and we will specialise to
this case when needed. We find that the critical Josephson current may
be modulated by changing the angle between the vectors describing
absence of inversion symmetry on each side. This effect is analogous
to tunnelling magnetoresistance in ferromagnetic tunnel
junctions.\cite{julliere_slonczewski} We also calculate the
quasiparticle current. For temperatures close to zero, the
current-voltage diagram may contain several discontinuities determined
by the relative size of the two gaps. We claim that both these results
may help to determine the possibly novel properties of the
superconducting state in materials like CePt$_3$Si, UIr and
Cd$_2$Re$_2$O$_7$.

The Hamiltonian considered is $H = H_{\mathrm{N}} + H_{\mathrm{SC}}$, where $H_{\mathrm{SC}}$ describes superconductivity. The normal state Hamiltonian is
\begin{equation}
  \label{eq:Hamnormal}
  H_{\mathrm{N}} = \sum_{\bk}  \phi^\dagger_{\bk} \left[\varepsilon_\bk + {\boldsymbol{B}}_\bk \cdot {\boldsymbol{\sigma}} \right] \phi_\bk,
\end{equation}
where $\phi^\dagger_\bk = (\cdgup, \, \cdgdn)$, $\varepsilon_\bk$ is
the band dispersion, and the vector $\boldsymbol{\sigma}$ consists of
the three Pauli matrices. We name the spin quantisation axis the
$z$-axis. The vector ${\boldsymbol{B}}_\bk$ removes the spin
degeneracy from the band $\varepsilon_\bk$. By a transformation to a
basis $\tilde{\phi}^\dagger_\bk = ( \cdgtd{+}, \, \cdgtd{-})$ where
\eqref{eq:Hamnormal} is diagonal, one finds $H_{\mathrm{N}} =
\sum_{\lambda=\pm,\bk}\epstd{\lambda} \cdgtd{\lambda} \ctd{\lambda}$.
The quasiparticle spectrum is $\epstd{\pm} = \varepsilon_\bk \pm
|{\boldsymbol{B}}_{\bk}|$. We define $B_{\bk,\pm} = B_{\bk,x} \pm \im
B_{\bk,y} = B_{\bk,\perp} \ex{\pm \im \varphi_{\bk}}$.

The vector ${\boldsymbol{B}}_\bk$ has the property
${\boldsymbol{B}}_{-\bk} = -{\boldsymbol{B}}_{\bk}$,\cite{frigeri}
where ${\boldsymbol{B}}_\bk$ characterises the absence of inversion
symmetry in the crystal. The origin may be that ions are removed from
high-symmetry positions, as in ferroelectrics,\cite{sergienko2}
leading to internal electric fields and thus increased spin-orbit
coupling.\cite{rashba,dresselhaus} To establish the form of
${\boldsymbol{B}}_\bk$, point group symmetry considerations may be
employed \cite{frigeri} and ${\boldsymbol{B}}_\bk$ will depend on the
direction in which the ions are displaced.

An electron with momentum $\bk$ will align its spin parallel or
antiparallel to ${\boldsymbol{B}}_\bk$. In a free electron model with
the Rashba interaction,\cite{rashba} the 1D density of states for the
$+$ and $-$ band at the Fermi level are equal.\cite{molenkamp} Still,
we allow these to be unequal, which is the general case.

Let us now turn to the term responsible for superconductivity,
$H_{\mathrm{SC}}$. We write down the interaction in terms of the
long-lived excitations in the normal state
\begin{equation}
  \label{eq:HamSC}
  H_{\mathrm{SC}} = \frac{1}{2} \sum_{\lambda\mu, \bk \bk'} V_{\lambda \mu}(\bk,\bk') \tilde{c}^\dagger_{\lambda,-\bk} \tilde{c}^\dagger_{\lambda,\bk} \tilde{c}_{\mu,\bk'} \tilde{c}_{\mu,-\bk'}.
\end{equation}
This model contains only intraband Cooper pairing. Interband Cooper
pairs are strongly suppressed if the spin-orbit splitting is much
larger than the superconducting gaps, even though the two bands may
touch at some isolated points on the Fermi surface.\cite{samokhin}
This is the limit we are investigating. Defining $\Delta_{\lambda,\bk}
= - \sum_{\mu,\bk'} V_{\lambda \mu} (\bk, \bk') \langle
\tilde{c}_{\mu,\bk'} \tilde{c}_{\mu,-\bk'} \rangle$, the standard mean
field approach gives the total Hamiltonian
\begin{equation}
  \label{eq:HamSCMF}
  H = \sum_{\lambda,\bk} \left[ \epstd{\lambda} \cdgtd{\lambda} \ctd{\lambda} + \frac{1}{2} \left(\Delta_{\lambda,\bk} \tilde{c}^\dagger_{\lambda,\bk} \tilde{c}^\dagger_{\lambda,-\bk} + \mathrm{h. c.}\right)\right].
\end{equation}
Note that $\Delta_{\lambda,-\bk} = - \Delta_{\lambda,\bk}$ follows
from the fermionic anticommutation relations. In Equation
\eqref{eq:HamSCMF}, the two bands are decoupled, resulting in Green's
functions diagonal in the band indices. This is a result of the mean
field approximation. $\Delta_{\pm,\bk}$ are in general not
independent, but related through the self-consistency equations due to
the possibility of interband pair scattering.\cite{mineev}

The relation ${\boldsymbol{B}}_{-\bk} = -{\boldsymbol{B}}_{\bk}$
ensures that states with opposite momenta within a band have opposite
spins. For a spin-1/2 state, the time reversal operator is ${\cal K} =
-\im \sigma_y {\cal K}_0$, where ${\cal K}_0$ denotes complex
conjugation. Let the original operators transform according to ${\cal
  K}: c^\dagger_{\bk,\sigma} = -\sigma c^\dagger_{-\bk,-\sigma}$ under
time reversal. The effect of time reversal on the new operators then
becomes ${\cal K}: \tilde{c}^\dagger_{\lambda,\bk} = t_{\lambda,\bk}
\tilde{c}^\dagger_{\lambda,-\bk}$, where $t_{\lambda,\bk} =
\ex{-\lambda \im \varphi_\bk}$. This means that if
$\chi_{\lambda,\bk}$ is the order parameter for pairs of time reversed
states, one finds $\Delta_{\lambda,\bk} = t_{\lambda,\bk}
\chi_{\lambda,\bk} $. This gives $\chi_{\lambda,\bk} =
\chi_{\lambda,-\bk}$. Thus, $\chi_{\lambda,\bk}$ may be expanded in
terms of even basis functions of irreducible representations of the
space group.\cite{sergienko}

Define the matrix $\Delta_\bk$ whose elements are the gap functions
$\Delta_{\bk,\sigma\sigma'}$ in the original basis where spin is
quantised along the $z$-axis. This may be written as
\begin{equation}
  \label{eq:gapOrig}
  \Delta_\bk =  \eta_{\bk,\mathrm{S}} \, g + \eta_{\bk,\mathrm{T}} ( \hat{{\boldsymbol{B}}}_\bk \cdot \boldsymbol{\sigma} ) \, g ,
\end{equation}
where $g = -\im \sigma_y$. The first term is symmetric in momentum
space and antisymmetric in spin space, whereas the opposite is the
case for the last term. Thus, in the absence of spatial inversion
symmetry, the order parameters in a spin basis have no definite
parity, but is in general a linear combination of singlet (S) and
triplet (T) parts.\cite{edelstein,gorkov,fujimoto_hayashi} The singlet
and triplet components are determined by $\eta_{\bk,\mathrm{S}} =
\left(\chi_{+,\bk} + \chi_{-,\bk} \right)/2$ and
$\eta_{\bk,\mathrm{T}} = \left(\chi_{+,\bk} - \chi_{-,\bk} \right)/2$,
respectively.  {\it This means that knowledge of $\chi_{\pm,\bk}$ and
  ${\boldsymbol{B}}_\bk$ could help determine the gap structure and
  the symmetry of the pairing state}. For non-centrosymmetric
materials like CePt$_3$Si, this is currently a matter of intense
study.\cite{frigeri,samokhin,sergienko,bonalde_yogi_izawa,fujimoto_hayashi}

The normal and anomalous Green's functions for each band are ${\cal
  G}_\lambda (\bk, \omega_n) = -(\im \omega_n +
\xi_{\lambda,\bk})/(\omega_n^2 + \xi_{\lambda,\bk}^2 +
|\chi_{\lambda,\bk}|^2)$ and ${\cal F}_\lambda (\bk, \omega_n) =
t_{\lambda,\bk} \chi_{\lambda,\bk}/(\omega_n^2 + \xi_{\lambda,\bk}^2 +
|\chi_{\lambda,\bk}|^2)$, respectively.  These are defined in the
standard way, see {\it e.g.} Ref. \onlinecite{sergienko}. $\omega_n$
is a fermion Matsubara frequency, $\xi_{\lambda,\bk} = \epstd{\lambda}
- \mu$ and $\mu$ is the chemical potential.

Consider tunnelling between two such superconductors, A and B. Let
system A be described by the Hamiltonian \eqref{eq:HamSCMF}. The
Hamiltonian of system B is defined equivalently, only with
$\cu,\ctd{\lambda} \rightarrow \dud, \tilde{d}_{\rho,\bp}$. Moreover,
we allow ${\boldsymbol{B}}_\bk^\A$ and ${\boldsymbol{B}}_\bp^\B$ to be
different. Consequently, even if $\bk = \bp$, the spin in a state $+$
or $-$ may be different on sides A and B. The tunnelling Hamiltonian
is $H_{\mathrm{T}} = \sum_{\lambda \rho, \bk \bp}
\left(\tilde{T}^{\lambda \rho}_{\bk \bp} \cdgtd{\lambda}
  \tilde{d}_{\rho,\bp} + \tilde{T}^{\lambda \rho \, \ast}_{\bk \bp}
  \tilde{d}_{\rho,\bp}^\dagger \ctd{\lambda} \right)$.  The tunnelling
matrix elements $\tilde{T}^{\lambda \rho}_{\bk \bp}$ depends strongly
on the {\it direction} of $\bk$ and $\bp$.\cite{bruder} Tunnelling is
much more probable for a momentum normal to the interface rather than
parallel to it.\cite{duke,harrison_scalapino,bruder} If we assume
that spin is conserved in the tunnelling process, {\it i.e.}
$H_{\mathrm{T}} = \sum_{\bk\bp,\sigma} T_{\bk \bp} \cdg \dud +
\mathrm{h. c.}$, we find $|\tilde{T}^{\lambda \rho}_{\bk \bp}|^2 =
|T_{\bk\bp}|^2( 1 + \lambda \rho \, \hat{{\boldsymbol{B}}}^\A_\bk
\cdot \hat{{\boldsymbol{B}}}^\B_\bp)/2$.

The number operator for band $\lambda$ in system A is $N^\A_\lambda =
\sum_\bk \cdgtd{\lambda} \ctd{\lambda}$.  We define
$\dot{N}^\T_\lambda = \im [H_\T, N^\A_\lambda]$, such that the charge
current is $I(t) = -e \sum_\lambda \langle \dot{N}^\T_\lambda
\rangle$. To lowest order in the tunnelling matrix elements, standard
theory gives $I(t) = I_{\mathrm{qp}} + I_{\mathrm{J}}(t)$, where
$I_{\mathrm{qp}} = -2e \sum_\lambda \mathrm{Im} \, \Phi_\lambda (eV)$
and $I_{\mathrm{J}}(t) = 2e \sum_\lambda \mathrm{Im}
\left[\Psi_\lambda(eV) \, \ex{2 \im eV t} \right]$.  The voltage is
$eV = \mu_\A - \mu_\B$. In the Matsubara formalism, we have
\begin{equation}
  \label{eq:PhiPsi}
  \begin{split}
   \Phi_\lambda (\omega_\nu) & = \frac{1}{\beta} \sum_{\genfrac{}{}{0cm}{1}{\bk\bp}{\rho,\omega_n}} |\tilde{T}^{\lambda \rho}_{\bk \bp}|^2 {\cal G}^\A_\lambda (\bk, \omega_n - \omega_\nu) {\cal G}^\B_\rho (\bp, \omega_n), \\
  \Psi_\lambda (\omega_\nu) & = \frac{1}{\beta} \sum_{\genfrac{}{}{0cm}{1}{\bk\bp}{\rho,\omega_n}} \tilde{T}^{\lambda \rho}_{\bk \bp} \tilde{T}^{\lambda \rho}_{-\bk,-\bp} {\cal F}^{\A \ast}_\lambda (\bk, \omega_n - \omega_\nu) {\cal F}^\B_\rho (\bp, \omega_n),
\end{split}
\raisetag{2cm}
\end{equation}
where $\omega_\nu \rightarrow eV + \im \, 0^+$ is a boson Matsubara
frequency. We have assumed that the bulk Green's functions may be
used, neglecting boundary effects. Such effects could however be of 
importance in these systems,\cite{yokoyama} due to the possibility of
subgap surface bound states or distortions of the order parameters close
to the surface.

Time-reversal symmetry of $H_\T$ gives $\tilde{T}^{\lambda
  \rho}_{-\bk,-\bp} = \tilde{T}^{\lambda \rho \, \ast}_{\bk \bp}
t^\A_{\lambda,\bk} t^{\B \ast}_{\rho,\bp}$.\cite{sergienko} These
phase factors will cancel the ones from the anomalous Green's
functions in Equation \eqref{eq:PhiPsi}, which shows that each band
$\lambda$ may behave as a singlet superconductor with gap function
$\chi_{\lambda,\bk}$.\cite{gorkov,sergienko}

We take the continuum limit \footnote{$\sum_\bk f_\lambda(\bk)
  \rightarrow \int \du \xi \int_{S(\xi)} \du \Omega \, {\cal
    N}_\lambda(\xi, \Omega) \, f_\lambda \left(\bk(\xi,\Omega)
  \right)$.  $S(\xi)$ is a constant energy surface in momentum space.
  ${\cal N}_\lambda(\xi, \Omega)$ is the density of states.} and
assume that ${\cal N}_\lambda(\xi, \Omega)$, the angle-resolved
density of states in band $\lambda$ in the non-superconducting phase,
is constant.

The gap $\chi_\lambda (\xi,\Omega)$ depends on both energy and the
direction of momentum. Neglecting the energy dependence is
standard.\cite{ambegaokar,werthamer} The tunnelling matrix elements
ensures that momenta approximately perpendicular to the interface will
dominate.\cite{duke,harrison_scalapino,bruder} We therefore let
$\chi_\lambda (\xi,\Omega)\approx \chi_\lambda$, the value at the
Fermi level and directions normal to the interface (remember that
$\chi_{\lambda,\bk}=\chi_{\lambda,-\bk}$). This is exact if the gaps
are isotropic or if the tunnelling is strictly one-dimensional. It
could also be a good approximation if the variations of $\chi_+$ and
$\chi_-$ are small in the region around normal incidence.  We define
$\chi_\lambda = |\chi_\lambda| \ex{\im \vartheta_\lambda}$.

The energy dependence of the tunnelling matrix elements may be
neglected. We will need the quantity $\tau_{\lambda \rho} = \int \du
\Omega^\A \int \du \Omega^\B |\tilde{T}^{\lambda \rho}(\Omega^\A ,
\Omega^\B)|^2$.  Let us look at a specific example, where $|T_{\bk
  \bp}|^2 \sim |T|^2 \hat{k}_\perp \hat{p}_\perp \Theta(\hat{k}_\perp
\hat{p}_\perp)$.\footnote{The Heaviside function $\Theta$ ensures that
  the momentum perpendicular to the interface does not change
  sign.\cite{bruder}} In addition, we choose the Rashba interaction
${\boldsymbol{B}}^\A_\bk = \alpha (\hat{\boldsymbol{n}}^\A \times
\bk)$.\cite{rashba} We let $\hat{\boldsymbol{n}}^\A$ and
$\hat{\boldsymbol{n}}^\B$, and consequently the nodes of
${\boldsymbol{B}}^\A_\bk$ and ${\boldsymbol{B}}^\B_\bp$, point
parallel to the interface. The Rashba interaction appears to be an
appropriate choice for CePt$_3$Si \cite{frigeri} and
Cd$_2$Re$_2$O$_7$.  \cite{sergienko} Define the angle $\zeta$ by $\cos
\zeta = \hat{\boldsymbol{n}}^\A \cdot \hat{\boldsymbol{n}}^\B$. This
gives
\begin{equation}
  \label{eq:exampleTunn}
  \tau_{\lambda \rho} = \frac{|T|^2}{2} \left(1 +  x \, \lambda \rho \, \cos \zeta \right),
\end{equation}
with $x \approx 0.6$.\footnote{We find $x = (2 \int_0^1 \du y
  \sqrt{1-y} K(y)/\pi)^2 \approx 0.6$, where $K(y)$ is the complete
  elliptic integral of the first kind.\cite{abramowitz}} Numerical
integration indicates that Equation \eqref{eq:exampleTunn} is a very
good approximation also when parallel momentum is conserved.
\footnote{$|T_{\bk \bp}|^2 \sim |T|^2 \hat{k}_\perp \delta_{\bk,\bp}$
  gives the same qualitative result. Refraction effects are not
  included, but this is negligible whenever $(|\bk_{\mathrm{F},-}| + |\bk_{\mathrm{F},+}|)/2
\gg ||\bk_{\mathrm{F},-}| - |\bk_{\mathrm{F},+}|| \sim \alpha$, at least in the important region,
  {\it i.e.}  close to normal incidence.} In general, it seems
reasonable that if mostly electrons near normal incidence contribute
to the current, Equation \eqref{eq:exampleTunn} is applicable with
$\cos \zeta \equiv \hat{{\boldsymbol{B}}}^\A_{{\boldsymbol{q}}_\perp}
\cdot \hat{{\boldsymbol{B}}}^\B_{{\boldsymbol{q}}_\perp}$ where
${\boldsymbol{q}}_\perp$ is perpendicular to the interface. $x \in
[0,1]$ is in fact an experimentally accessible quantity. This will be
discussed below.

The conductance in the normal phase is given by $G_{\mathrm{N}} \equiv
I_{\mathrm{N}}/V = 2 e^2 \pi \sum_{\lambda \rho} {\cal N}^\A_\lambda
{\cal N}^\B_\rho \, \tau_{\lambda \rho}$. Define $r_{\mathrm{N}}
(\zeta) \equiv G_{\mathrm{N}}(\zeta)/G_{\mathrm{N}}(\pi/2)$. Using
Equation \eqref{eq:exampleTunn}, we find $r_{\mathrm{N}} (\zeta) = 1 +
(1-d)^2/(1+d)^2 \, x \cos \zeta$ where $d \equiv {\cal N}_-/{\cal
  N}_+$ is the ratio of the densities of states. The dependence on the
angle $\zeta$ is similar to tunnelling magnetoresistance between
ferromagnets \cite{julliere_slonczewski} and vanishes if ${\cal N}_+ =
{\cal N}_-$.

In the superconducting phase, the quasiparticle current for $T
\rightarrow 0$ becomes $I_{\mathrm{qp}} = \pi e \sum_{\lambda \rho}
{\cal N}^\A_\lambda {\cal N}^\B_\rho \, \tau_{\lambda \rho} \,
\gamma^{\A \B}_{\lambda \rho}$, where
\begin{align}
  \label{eq:gammaliten}
  \gamma^{\A \B}_{\lambda \rho} & =  \Theta \left(|eV| - (|\chi^\A_\lambda| + |\chi^\B_\rho|)\right) \, eV \, \sqrt{1-\delta_{\lambda \rho}^2} \\
   & \quad \times \left[2 E \left(\frac{1-\sigma_{\lambda \rho}^2}{1-\delta_{\lambda \rho}^2} \right) - \frac{\sigma_{\lambda \rho}^2-\delta_{\lambda \rho}^2}{1-\delta_{\lambda \rho}^2} K \left(\frac{1-\sigma_{\lambda \rho}^2}{1-\delta_{\lambda \rho}^2} \right) \right].  \nonumber
\end{align}
$K(m)$ and $E(m)$ are the complete elliptic integrals of the first and
second kind, respectively.\cite{abramowitz} We have defined
$\sigma_{\lambda \rho} = (|\chi^\A_\lambda| + |\chi^\B_\rho|)/eV$ and
$\delta_{\lambda \rho} = (|\chi^\A_\lambda| - |\chi^\B_\rho|)/eV$.
This is a two-band generalisation of the one-band $s$-wave expression.\cite{werthamer} {\it The usual one-band treshold at $eV =
  \chi^\A + \chi^\B$ is replaced by at most four discontinuities}.

At zero voltage difference, the Josephson current becomes
$I_{\mathrm{J}} = 4 \pi e \sum_{\lambda \rho} {\cal N}^\A_\lambda
{\cal N}^\B_\rho \, \tau_{\lambda \rho} \, \Gamma^{\A \B}_{\lambda
  \rho} \, \sin (\vartheta^\B_\rho - \vartheta^\A_\lambda)$, where
\begin{equation}
  \label{eq:Gamma}
  \Gamma^{\A \B}_{\lambda \rho} = \frac{|\chi^\A_\lambda||\chi^\B_\rho|}{|\chi^\A_\lambda| + |\chi^\B_\rho|} \, K\left(\frac{(|\chi^\A_\lambda| - |\chi^\B_\rho|)^2}{(|\chi^\A_\lambda| + |\chi^\B_\rho|)^2}  \right).
\end{equation}
This is a general two-band $s$-wave expression valid when interband Cooper
pairs are absent. It is a straightforward generalisation of the
standard one-band result.\cite{ambegaokar,werthamer}

We now consider the case of spin-orbit split bands. Usually, for equal
systems one could let $|\chi^\A_\pm| = |\chi^\B_\pm| \equiv
|\chi_\pm|$. This might not always be justified in our model, since
the direction dependence of the gaps might depend on the nature of
${\boldsymbol{B}}^{\A (\B)}_{\bk (\bp)}$. However, let us again turn
to the example above, where $\hat{\boldsymbol{n}}^\A$ and
$\hat{\boldsymbol{n}}^\B$ point parallel to the interface. This makes
$\bk,\bp \sim {\boldsymbol{q}}_\perp$ equivalent directions even
though $\hat{\boldsymbol{n}}^\A \neq \hat{\boldsymbol{n}}^\B$, at
least in the isotropic approximation. The above assumption should then
be justified and will be used below. In addition, since the model
\eqref{eq:HamSC} contains interband pair scattering, we consider
phase-locked bands where $\vartheta_+ = \vartheta_- + n \pi$ and $n$
is zero or one. We do not investigate the possibility of small
oscillations of the interband phase difference.\cite{leggett}

Whenever $|\chi_+| = |\chi_-|$ and ${\cal N}_+ = {\cal N}_-$,
$I_{\mathrm{qp}}$ becomes independent of $\zeta$ and equals the
one-band result. In the case of unequal gaps, extra discontinuities
should appear in the current-voltage characteristics.
\begin{figure}
  \centering
  \scalebox{0.45}{
    \hspace{-1cm}
   \includegraphics{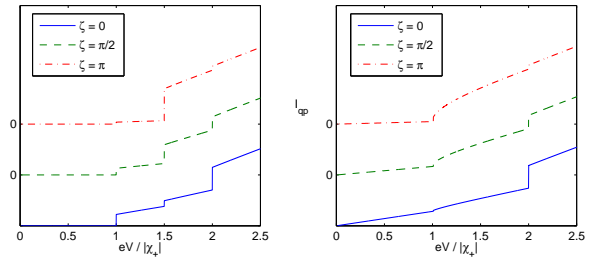}
  }
  \caption{(Color online) The quasiparticle current-voltage diagram at
    $T=0$ for angles $\zeta = 0,\pi/2,\pi$, $x=0.7$ and $d=1$. The
    graphs are displaced in the vertical direction for clarity. {\it
      Left panel}: $|\chi_-|/|\chi_+| = 0.5$. {\it Right panel}:
    $|\chi_-|/|\chi_+| = 0.01$. }
\label{fig:double}
\end{figure}
Figure \ref{fig:double} shows the quasiparticle current at $T=0$ as
function of voltage for the three angles $\zeta = 0$, $\pi/2$ and
$\pi$ with $x=0.7$ and $d=1$. The graphs are displaced in the vertical
direction for clarity. Note that as $x \rightarrow 0$ all cases
approach the $\zeta=\pi/2$ graph. In the left panel,
$|\chi_-|/|\chi_+| = 0.5$. Discontinuities appear at $eV = 2
|\chi_-|$, $(|\chi_+| + |\chi_-|)$ and $2 |\chi_+|$. In the right
panel, where $|\chi_-|/|\chi_+| = 0.01$, current can flow also for
small voltages, although this almost vanishes as $\zeta \rightarrow
\pi$. Smearing of the steps due to interband scattering of
quasiparticles should be negligible as long as the gap difference is
well above $k_B T$. This breaks down in the limit of equal gaps, but
then the current-voltage diagram collapse to the one-band result
anyway. Anisotropic gaps, gap nodes and non-zero temperature
may in general also lead to smearing. At non-zero temperatures,
logarithmic singularities in $I_{\mathrm{qp}}$ at $eV = \left||\chi_+|
  - |\chi_-|\right|$ may show up.\cite{barone}

Define $|\chi_\mathrm{M}| = \mathrm{max}(|\chi_+|,|\chi_-|)$ and
$|\chi_\mathrm{m}| = \mathrm{min}(|\chi_+|,|\chi_-|)$. The ratio $F
\equiv |\chi_\mathrm{m}|/|\chi_\mathrm{M}|$ is easily found from the
position of the first and last discontinuity. In addition, if
$D_{\lambda \rho}$ denotes the jump in $I_{\mathrm{qp}}$ at $eV =
(|\chi_\lambda|+|\chi_\rho|)$, one finds $d^2 F =
D_\mathrm{{mm}}/D_\mathrm{{MM}}$, where $d \equiv {\cal
  N}_\mathrm{m}/{\cal N}_\mathrm{M}$. These methods for finding $F$
and $d^2 F$ are independent of $\zeta$ and $x$. Furthermore, $x$ may
be determined from $x \cos \zeta = (2 \sqrt{D_\mathrm{{mm}}
  D_\mathrm{{MM}}} - D_{+-})/(2 \sqrt{D_\mathrm{{mm}} D_\mathrm{{MM}}}
+ D_{+-})$, for any angle $\zeta$.

The critical Josephson current $I_\mathrm{J,c}$ will also depend on
$\zeta$. There is a close analogy to tunnelling
magnetoresistance.\cite{julliere_slonczewski} Define $a = (-1)^n 8 \,
{\cal N}_\mathrm{+} {\cal N}_\mathrm{-} \Gamma_{+-} /\pi({\cal
  N}_\mathrm{+}^2 |\chi_+|+{\cal N}_\mathrm{-}^2 |\chi_-|)$, where
$|a| = |a|(F,d)$ is monotonically increasing for $F \leq 1$,
$|a|(0,d)=0$ and $|a|(1,d)=2d/(1+d^2) \leq 1$. We find that $a = -1$
results in $I_\mathrm{J,c} (\pi/2) = 0$. For $a \neq -1$, we define
\begin{equation}
  \label{eq:ratiodef}
  r_\mathrm{J}(\zeta) \equiv \frac{I_\mathrm{J,c} (\zeta)}{I_\mathrm{J,c} \left(\frac{\pi}{2}\right)} = \Bigg|1 + \frac{1-a}{1+a} \, x \cos \zeta \Bigg|,
\end{equation}
showing the possible modulation of the critical Josephson current with
$\zeta$. In addition, $a = (1-r_\mathrm{J}(0) + x)/(r_\mathrm{J}(0) -
1 + x)$. The sign of $a$ determines $n$ and hence the relative sign
between $\chi_{+,{\boldsymbol{q}}_\perp}$ and
$\chi_{-,{\boldsymbol{q}}_\perp}$. {\it One may then determine the
  ratio between the singlet and triplet components of the order
  parameters in a spin basis, since
  $|\eta_{{\boldsymbol{q}}_\perp,\mathrm{S}}|/|\eta_{{\boldsymbol{q}}_\perp,\mathrm{T}}|
  = (1+(-1)^n F)/(1-(-1)^n F)$}.

Note that if $|\chi_+| = |\chi_-|$ and $n=0$, $r_\mathrm{J}(\zeta) =
r_\mathrm{N}(\zeta)$. If in addition $d=1$, $I_\mathrm{J,c}$ becomes
independent of $\zeta$ and the one-band result \cite{ambegaokar} is
recovered. A modulation may be a result of unequal gaps ($F \neq 1$),
unequal densities of states ($d \neq 1$) or both. In addition, no
modulation of $r_\mathrm{J}(\zeta)$ could be interpreted both as $x=0$
and $a=1$. Both these ambiguities should be distinguishable through
the quasiparticle current-voltage characteristics. Consistency demands
that $|a|(F,d)$ found from the Josephson current fits $F$ and $d$
found from the quasiparticle current.

To determine that the jumps in the quasiparticle current arise from
spin-orbit split bands due to breakdown of inversion symmetry along a
certain axis, several junctions with different relative orientations
of those axes would be needed.\footnote{In our model, this amounts to
  controlling the angle between $\hat{\boldsymbol{n}}^\A$ and
  $\hat{\boldsymbol{n}}^\B$.} The synthesis and manipulation of such
junctions thus represents a considerable experimental challenge.
However, building Josephson junctions with controllable
crystallographic orientations was essential to proving the $d$-wave
symmetry of the order parameter in the high-$T_c$
cuprates.\cite{tsuei} Also, the presence of two gaps may be possible
to detect in other experiments, such as scanning tunnelling
microscopy.

In conclusion, we predict possible new effects in the tunnelling
current between non-centrosymmetric superconductors with significant
spin-orbit splitting. Spin conservation in the tunnelling barrier may
then result in a modulation of the critical Josephson current when
varying the relative angle between the vectors describing absence of
inversion symmetry on each side. We have also shown that several
discontinuities may appear in the quasiparticle current. We have
argued that both these phenomena might help to determine the possibly
exotic gap symmetry and pairing state of non-centrosymmetric
superconductors.

We thank Thomas Tybell for valuable discussions. This work was
supported by the Research Council of Norway, Grant Nos. 158518/431,
158547/431, and 167498/V30 (NANOMAT).

\end{document}